\documentstyle[a4,
twoside]{article}
\makeatletter
\renewcommand{\@oddhead}{An Interaction of ...
 . III. Linear Friction as Radiation Reaction  \hfill \thepage}
\renewcommand{\@evenhead}{\thepage \hfill S.A. Choro\v{s}avin }
\renewcommand{\@oddfoot}{}
\renewcommand{\@evenfoot}{}
\makeatother

\author{S.A.~Choro\v{s}avin}
\title{ 
  1D Particle, 1D Field, 1D Interaction.
 \bigskip\\
{\large  Simple Exactly Solvable Models based on \\ 
  Finite Rank Perturbations Methods. }
 \bigskip\\ 
    III. Linear Friction as Radiation Reaction 
 }
\date{}

\begin{document}
\maketitle 
\begin{abstract}
 This paper is an electronic application to my set of lectures, 
 subject:`Formal methods in solving differential equations and 
 constructing models of physical phenomena'. 
 Addressed, mainly: postgraduates and related readers. 
 Content: a discussion of the simple models of linear friction, 
 the models, that have the mechanism that is based on radiation reaction. 
 The interactions we will deal are based 
 on equation arrays of the kind: 
\begin{eqnarray*}
 \frac{\partial^2 q(t)}{\partial t^2}
 &=&-\Omega^2 q(t)+f_{compl}(t,q,Q) 
\\
 \frac{\partial^2 u(t,x)}{\partial t^2} 
 &=&
 c^2\frac{\partial^2 u(t,x)}{\partial x^2}
 -4{\gamma }c\delta(x-x_0)\Big(F_{src}(t,q,Q)\Big)+f_1(t,x) 
\\
 Q(t)  
 &=& <l(t)|u> 
\\
\end{eqnarray*}
 Central mathematical points: d'Alembert-Kirchhoff-like formulae. 
 Central physical points: phenomena of Radiation Reaction, Braking Radiation 
 and Friction.
\end{abstract}

\newpage 
\section*%
{ Introduction. } 
\subsection*%
{ A Harmonic Oscillator Coupled to an One-Dimensional Field. }

 We will discuss a possible description of a detail of dynamical behaviour of  
 one-dimensional newtonian particle. The detail we are here interested in 
 is named `friction'.  
 We will focus on only two aspects of this phenomenon: 
 First, if a particle MOVES through a medium, e.g., 
 through water, then a special force arises, the force which acts on 
 the particle so that it brakes the particle's moving, 
 "tries to stop" the particle, 
 and the particle's energy decreases. 
 At the same time, a special medium motion arises: or medium waves 
 of this or that or other kind arise, or the medium becomes more warm, 
 or the medium generates a light... 
 In such cases as these, 
 one associates these two phenomena,
 thinking of them as reciprocal ones, 
 conceiving them as the result of interplay, 
 and says the particle generates
\footnote{ or, 'is a source of'}
 a kind of radiation which brakes 
 the particle's motion; 
 so, one says about the {\em braking or damping radiation}.

 I am now trying to express this point 
 in a language of mathematical formulae, 
 --- it is just the subject of this paper. 

 \addvspace{\bigskipamount}

 The Newtonian equation of motion of the particle, that moves under the action 
 of an external force, 
$F_{external}(t,q,\cdots)$, 
 reads: 
 \begin{eqnarray*}
 M
 \frac{\partial^2 q(t)}{\partial t^2}
 &=&F_{external}(t,q,\cdots) 
 \end{eqnarray*}
 where 
$M$
 stands for the mass of the particle. 
 For example, 
 the Newtonian equation of motion of the linear harmonic oscillator, 
 which moves being subjected to an external complementary
\footnote{ external to the oscillator, as a physical system}
 force 
$F_{ext,osc}(t,q,\cdots)$, 
 is:
\begin{eqnarray*}
 M
 \frac{\partial^2 q(t)}{\partial t^2}
 &=&
 -k q(t)+F_{ext,osc}(t,q,\cdots) 
\end{eqnarray*}
 This equation is often written as 
\begin{eqnarray*}
 \frac{\partial^2 q(t)}{\partial t^2}
 &=&
 -\Omega^2 q(t)+f_{compl}(t,q,\cdots) 
\end{eqnarray*}
 where 
$$
 \Omega^2 := k/M\,,\, f_{compl}(t,q,\cdots):=F_{ext,osc}(t,q,\cdots)/M \,. 
$$
 We will deal with the case, where 
$$
 F_{external}(t,q,\cdots)
 = -2{\gamma_v}\frac{\partial q(t)}{\partial t}+F_{compl}(t,q,\cdots) 
\,, 
$$
 that is, 
\begin{eqnarray*}
 M
 \frac{\partial^2 q(t)}{\partial t^2}
 &=&-2{\gamma_v}\frac{\partial q(t)}{\partial t}+F_{compl}(t,q,\cdots) 
\end{eqnarray*}
 In particular, if we deal with the linear harmonic oscillator, then 
$$
 f_{compl}(t,q,\cdots)
 = -2{\gamma}\frac{\partial q(t)}{\partial t}+f_{compl,0}(t,q,\cdots)
$$
  where 
$$
 {\gamma} := {\gamma_v}/M \,,
$$
 that is,
\begin{eqnarray*}
\frac{\partial^2 q(t)}{\partial t^2}
 &=&-2{\gamma}\frac{\partial q(t)}{\partial t}
 -\Omega^2 q(t)+f_{compl,0}(t,q,\cdots) \,,
\end{eqnarray*}
 It is just
$-2{\gamma_v}\frac{\partial q(t)}{\partial t}$, 
 the term, by means of which one simulates the physical effect 
 that one is used to naming {\em ``linear''} friction. 
 Now then, I want to interpret the appearence of this term as an effect 
 of radiation reaction, an effect of an interaction of the particle with 
 a field.
 So, I have to declare models of the fields and models of the interactions.

 In this paper I will discuss several models of 
 one-dimensional particle 
 coupled to one-dimensional scalar field. 
 Primarily I am interested in the model described by the equation array
\footnote{
$4{\gamma }c\rho = k\,, \Omega^2 = k/M$ ;  
 the constants of the model mean, e.g.:
$c=$ 
 propagating waves velocity, 
$\rho = $ 
 `a density' of the field, 
$k = $ 
 elasticity constant, 
$M = $ 
 the mass of the particle.
 Of course, we assume 
$c>0$ . 
}
\begin{eqnarray*}
 \frac{\partial^2 q(t)}{\partial t^2}
 &=&-\Omega^2\Big(q(t)-Q(t)\Big)+f_0(t) 
\\
 \frac{\partial^2 u(t,x)}{\partial t^2} 
 &=&
 c^{2}\frac{\partial^2 u(t,x)}{\partial x^2}
 -4{\gamma }c\delta(x-x_0)\Big(Q(t)-q(t)\Big)+f_1(t,x) 
\\
 Q(t)  
 &=& u(t,x_0) 
\\
\end{eqnarray*}
 I am also interested in the model described by the equation array
\begin{eqnarray*}
 \frac{\partial^2 q(t)}{\partial t^2}
 &=&-\Omega^2 q(t) +\gamma_1 Q(t) +f_0(t) 
\\
 \frac{\partial^2 u(t,x)}{\partial t^2} 
 &=&
 c^{2}\frac{\partial^2 u(t,x)}{\partial x^2}
 -4{\gamma }c\delta(x-x_0)\Big(\frac{\partial q(t)}{\partial t}\Big)+f_1(t,x) 
\\
 Q(t)  
 &=& \alpha_1\frac{\partial u(t,x_0)}{\partial t} 
\\
\end{eqnarray*}
 One can say that these are models of a point interaction.
 In this paper I will also discuss several modification 
 of these models.
 All they are described by an equation array of the form 
\begin{eqnarray*}
 \frac{\partial^2 q(t)}{\partial t^2}
 &=&-\Omega^2 q(t)+f_{compl}(t,q,Q) 
\\
 \frac{\partial^2 u(t,x)}{\partial t^2} 
 &=&
 c^2\frac{\partial^2 u(t,x)}{\partial x^2}
 -4{\gamma }c\delta(x-x_0)\Big(F_{src}(t,q,Q)\Big)+f_1(t,x) 
\\
 Q(t)  
 &=& <l(t)|u> 
\\
\end{eqnarray*}
 where 
$l(t)$
 stands for a functional
\footnote{ 
 I use a P.A.M. Dirac's ``bra-ket'' syntax and suppose that 
$q$ and $Q$
 are usual (one-dimensional) functions of 
$t$ : 
$$
  q=q(t) \,,\quad Q=Q(t) \,, 
$$ 
 }
 , for any 
$t$
 fixed; in this formula we consider 
$t$ 
 as a free variable 
\footnote{ 
 thus, we deal with a family 
$\{l(t)\}_{t}$ 
 of functionals 
 ; 
 we will normally suppose that every 
$l(t)$
 is linear, for any 
$t$
 fixed. 
 Moreover, we will deal with the case where 
$<l(t)|u>$
 is of the form 
$$
 <l(t)|u> 
 := \alpha_0 u(t,x_0)+\alpha_1 \frac{\partial}{\partial t}u(t,x_0)
$$ 
 }
 .
 After indicating {\bf d'Alembert-Kirchhoff-like formulae }
 for solutions to these systems I obtain insulated effective 
 equations of motion of 
$q(t)$ 
 and then I briefly compare them.

 A few words about the THREE-dimensional particle: 
 the case is very complicated, however we can formally reduce it to the case 
 where ONE THREE-dimensional particle interacts with THREE ONE-dimensional 
 scalar fields, or, ONE ONE-dimensional VECTOR field, e.g., 
\begin{eqnarray*}
 \frac{\partial^2 q_i(t)}{\partial t^2}
 &=&-\Omega^2 q_i(t) +\gamma_1 Q_i(t) +f_{0,i}(t) 
\\
 \frac{\partial^2 u_i(t,x)}{\partial t^2} 
 &=&
 c^{2}\frac{\partial^2 u_i(t,x)}{\partial x^2}
 -4{\gamma }c\delta(x-x_0)\Big(\frac{\partial q_i(t)}{\partial t}\Big)
 +f_{1,i}(t,x) 
\\
 Q_i(t)  
 &=& \alpha_1\frac{\partial u_i(t,x_0)}{\partial t} 
 \qquad (\mbox{ here } i=1,2,3)
\\
\end{eqnarray*}

 It's all that I want here to say on this difficult topic... 

\newpage\section
{ Models of a Point Interaction of an only one-dimensional Oscillator with 
  an only one-dimensional Scalar Field }

 In this paper we fix measure units 
 and let
$x$ 
 be dimensionless position parameter, i.e., 
$$
 \mbox{ physical position coordinate } 
  =  [\mbox{ length unit }]\times x +const \,. 
$$
 Otherwise a confusion can ocurr, in relating to the definition 
$$
 \int_{-\infty}^{\infty}\delta(x-x_0)f(x)dx=f(x_0) \,.
$$
 We assume the standard foramalism, where 
$$ 
 \delta(x-x_0) = \frac{\partial 1_{+}(x-x_0)}{\partial x} 
$$
 and where 
$1_{+}$
 stands for a unit step function (Heaviside function):
$$ 
 1_{+}(\xi) = \Bigl\{ 
 \begin{array}{ccc}
 1&,&\mbox{ if } \xi \geq 0 \,,
 \\
 0&,&\mbox{ if } \xi < 0 \,,
 \\ 
 \end{array}
$$ 
\subsection
{ D'Alembert-Kirchhoff-like formulae }

 Recall that standard D'Alembert-Kirchhoff formulae read: 
 if 
$$
 \frac{\partial^2 u}{\partial t^2}
  = c^2\frac{\partial^2 u}{\partial x^2} + f
 \,,\quad 
 u=u(t,x) 
 \,,\quad 
 f=f(t,x) 
 \,,\quad 
 (t\geq 0)
           \eqno(*)
$$
 and given initial data, 
$u(0,\cdot )$ 
 and 
$\frac{\partial u(t,\xi)}{\partial t})\Big|_{t=0}$,
 then 
\begin{eqnarray*}
 u=u(t,x)&=&
 \frac{1}{2c}
 \int_0^t 
 \Big( \tilde f(\tau,x+c(t-\tau))-\tilde f(\tau,x-c(t-\tau))\Big) d\tau 
 + u_{0}(t,x)
\end{eqnarray*}
\begin{eqnarray*}
 u_{0}(t,x)
 &=& 
 c_{+}(x+ct) + c_{-}(x-ct)
 \\&=&
 \frac12 \Big(u(0,x+ct) + u(0,x-ct)\Big) 
   + \frac{1}{2c}\int_{x-ct}^{x+ct}
 (\frac{\partial u(t,\xi)}{\partial t})\Big|_{t=0} d\xi 
\end{eqnarray*}
 and where 
$\tilde f$ 
 stands for any Primitive function of  
$f$, 
 i.e., 
$$
  \frac{\partial\tilde f(t,x)}{\partial x} = f(t,x)
$$
 Note that
$$
 \tilde f(\tau,x+c(t-\tau))-\tilde f(\tau,x-c(t-\tau))
$$
 does not depend on what a primitive is one has chosen!!! 
 Moreover, we need only  
$ \tilde f $
 and not 
$f$
 itself.

 Now, I specify 
$f$.
 In this paper I will take  
$$
  f=-4{\gamma }c\delta(x-x_0)\Big(F_{src}(t)\Big)+f_1(t,x) 
$$
 For such an  
$f$  
 I conclude that 
$$
 \tilde f=-4{\gamma }c1_{+}(x-x_0)\Big(F_{src}(t)\Big)+\tilde f_1(t,x), 
$$
 and then I infer that 
\begin{eqnarray*}
 u(t,x)
 &=&
 -{2\gamma}
 \int_0^t \Big( 1_{+}(x+c(t-\tau)-x_0)-1_{+}(x-c(t-\tau)-x_0)\Big)
 \Big(F_{src}(\tau)\Big)d\tau
 \\&&{} + \frac{1}{2c}\int_0^t 
   \Big(\tilde f_1(\tau,x+c(t-\tau))-\tilde f_1(\tau,x-c(t-\tau))\Big) d\tau 
 +  u_{0}(t,x)
\end{eqnarray*}
 Denote now, to be more concise, 
$$ 
 u_{01}(t,x)
 := 
  \frac{1}{2c}\int_0^t 
  \Big(\tilde f_1(\tau,x+c(t-\tau))-\tilde f_1(\tau,x-c(t-\tau))\Big) d\tau 
  +  u_{0}(t,x)
$$ 
 and then rewrite the recent relation 
 as following: 
\begin{eqnarray*}
 u(t,x)
 &=&
 -{2\gamma}
 \int_0^t \Big( 1_{+}(x+c(t-\tau)-x_0)-1_{+}(x-c(t-\tau)-x_0)\Big)
 \Big(F_{src}(\tau)\Big)d\tau
 \\&&{} + u_{01}(t,x) 
\end{eqnarray*}

 Let us now analyse this expression. 
 We have: if 
$
 c\tau \not= ct+(x-x_0) 
$
, and if 
$
 c\tau \not= ct-(x-x_0) 
$
, then 
\begin{eqnarray*}
\makebox[5ex][l]{$\displaystyle 
 1_{+}(x+c(t-\tau)-x_0)-1_{+}(x-c(t-\tau)-x_0)
  $}
\\ &=&
 \Big\{ 
 \begin{array}{rcl}
  1 & , & c\tau < ct+(x-x_0) \\ 
  0 & , & ct+(x-x_0) < c\tau \\ 
 \end{array}
 - 
 \Big\{ 
 \begin{array}{rcl}
  1 & , & ct-(x-x_0) < \tau \\ 
  0 & , & c\tau < ct-(x-x_0) \\ 
 \end{array}
\\ &=&
 \Bigg\{ 
 \begin{array}{rcl}
  1 & , & c\tau < ct-|x-x_0| \\ 
  0 & , & ct-|x-x_0| < c\tau < ct+|x-x_0| \\ 
 -1 & , & ct+|x-x_0| < c\tau \\ 
 \end{array}
\\ &=&
 \Bigg\{ 
 \begin{array}{rcl}
  1 & , & \tau < t-|x-x_0|/c \\ 
  0 & , & t-|x-x_0|/c < \tau < t+|x-x_0|/c \\ 
 -1 & , & t+|x-x_0|/c < \tau \\ 
 \end{array}
\end{eqnarray*}
 Hence, for 
$t\geq 0$, 
$$
\parbox{\textwidth}{
\begin{eqnarray*}
 u(t,x) 
 &=&
 -{2\gamma}
 \int_0^{t-|x-x_0|/c} 
 \Big(F_{src}(\tau)\Big)d\tau \cdot 1_{+}(t-|x-x_0|/c)
 \\[\medskipamount]\\
 &&{}\qquad + u_{01}(t,x) 
 \\[\smallskipamount] \\ 
\end{eqnarray*}
}
$$
%
%
%
%
%
 Finally, this point of the analysis has an interesting consequence: 
\begin{eqnarray*}
 u(t,x_0) - u_{01}(t,x_0)
 &=&
 -{2\gamma}
 \int_0^t 
 \Big(F_{src}(\tau)\Big)d\tau 
\end{eqnarray*}
 and  
$$
\parbox{\textwidth}{
\begin{eqnarray*}
 u(t,x) 
 &=&
 \left\{ 
 \begin{array}{ccl}
 u(t-|x-x_0|/c,x_0) - u_{01}(t-|x-x_0|/c,x_0)
                                &,& \mbox{ if } 0 \leq t-|x-x_0|/c \\ 
                 0              &,& \mbox{ if } t-|x-x_0|/c < 0 \leq t \\ 
 \end{array}
 \right\} 
 \\[\medskipamount]\\
 &&{}\qquad + u_{01}(t,x) 
 \\[\smallskipamount] \\ 
\end{eqnarray*}
}
$$

 \newpage 

\subsection
{ Oscillator interacting with a scalar field }

 Recall that a standard relation which one is used to describing 
 one-dimensional harmonic oscillator 
 subjected to an external complementary force 
$F_{ext,osc}(t)$ 
 is this: 
$$
 \frac{\partial^2 q(t)}{\partial t^2}
  =-\Omega^2 q(t)+f_{compl}(t) 
$$
$$
 f_{compl}(t)=F_{ext,osc}(t)/M 
 \,,\, 
 M= \mbox{ the mass of the oscillated particle }
$$
 Recall also, in this paper I will discuss several systems described by

\begin{eqnarray*}
 \frac{\partial^2 q(t)}{\partial t^2}
 &=&-\Omega^2 q(t)+f_{compl}(t,q,Q) 
\\
 \frac{\partial^2 u(t,x)}{\partial t^2} 
 &=&
 c^2\frac{\partial^2 u(t,x)}{\partial x^2}
 -4{\gamma }c\delta(x-x_0)\Big(F_{src}(t,q,Q)\Big)+f_1(t,x) 
\\
 Q(t)  
 &=& <l(t)|u> 
\\
\end{eqnarray*}
 where 
$l(t)$
 stands for a functional, for any 
$t$
 fixed; in this formula we consider 
$t$ 
 as a free variable 
\footnote{ 
 thus, we deal with a family 
$\{l(t)\}_{t}$ 
 of functionals 
 ; 
 we will normally suppose that 
$l(t)$
 is linear, for any 
$t$
 fixed. 
 Moreover, we will deal with the case where 
$<l(t)|u>$
 is of the form 
$$
 <l(t)|u> 
 := \alpha_0 u(t,x_0)+\alpha_1 \frac{\partial}{\partial t}u(t,x_0)
$$ 
 }
 .
 In this case 
 I rewrite the recent d'Alembert-Kirchhoff relation 
 as following: 
\begin{eqnarray*}
 u(t,x)
 &=&
 -{2\gamma}
 \int_0^{t-|x-x_0|/c} 
 \Big(F_{src}(\tau,q,Q)\Big)d\tau \cdot 1_{+}(t-|x-x_a|/c)
 \\&&{} + u_{01}(t,x) 
\end{eqnarray*}
 I have now seen: given
$q$
 and 
$u_{01}$,
 then, 
\bigskip 

\fbox{\bf 
 in order to obtain 
$ u(t,x) $
 I need to obtain ONLY 
$ Q(t) \equiv <l(t)|u> $ } 
\bigskip 
\\
 After this observation use the last formula for 
$u(t,x)$
  and then obtain
\begin{eqnarray*}
 Q(t)
 &=&\Big<l(t)\Big|
 \Bigg\{ 
 -{2\gamma}
 \int_0^{t-|x-x_0|/c} 
 \Big(F_{src}(\tau,q,Q)\Big)d\tau \cdot 1_{+}(t-|x-x_0|/c)
 \\&&{} + u_{01}(t,x) \Bigg\}_{(
 \mbox{ 
 we have to consider this expression as a function of  $t, x$ }
 )} 
 \Big> 
\end{eqnarray*}

 We restrict ourselves to the case, where  
$$
 <l(t)|u> 
 := \alpha_0 u(t,x_0)+\alpha_1 \frac{\partial}{\partial t}u(t,x_0)
$$ 
 i.e., 
\begin{eqnarray*}
 Q(t)
 &=&
 \Big(\alpha_0 + \alpha_1 \frac{\partial}{\partial t} \Big)
 \Bigg(
 -{2\gamma}
 \int_0^{t-|x_0-x_0|/c} 
 \Big(F_{src}(\tau,q,Q)\Big)d\tau \cdot 1_{+}(t-|x_0-x_0|/c)
 \\&&{} + u_{01}(t,x_0) \Bigg)\Bigg|_{x=x_0}  
\end{eqnarray*}
 Hence 

\begin{eqnarray*}
 Q(t)
 &=&
 \Big(\alpha_0 + \alpha_1 \frac{\partial}{\partial t} \Big)
 \Bigg(
 -{2\gamma}
 \int_0^{t} 
 \Big(F_{src}(\tau,q,Q)\Big)d\tau 
 + u_{01}(t,x_0) \Bigg)  
\end{eqnarray*}
 and we have obtained: 

\begin{eqnarray*}
 \frac{\partial^2 q(t)}{\partial t^2}
 &=&-\Omega^2 q(t)+f_{compl}(t,q,Q) 
\\
 Q(t)
 &=&
 \Big(\alpha_0 + \alpha_1 \frac{\partial}{\partial t} \Big)
 \Bigg(
 -{2\gamma}
 \int_0^t 
 \Big(F_{src}(\tau,q,Q)\Big)d\tau
 + u_{01}(t,x_0) \Bigg)  
\\
 Q(t)  
 &=&
 \Big(\alpha_0 + \alpha_1 \frac{\partial}{\partial t} \Big)
 \Bigg\{ u(t,x) \Bigg\} \Bigg|_{x=x_0}  
\\
\end{eqnarray*}

\begin{eqnarray*}
 u(t,x) 
 &=&
 \displaystyle
 -{2\gamma}
 \int_0^{t-|x-x_0|/c} 
 \Big(F_{src}(\tau)\Big)d\tau \cdot 1_{+}(t-|x-x_a|/c)
 \\[\medskipamount]\\
 &&{}\qquad + u_{01}(t,x) \qquad \mbox{ , if } t\geq 0 \,.
 \\[\smallskipamount] \\ 
\end{eqnarray*}
 The specific 
$F_{src}$ 
 and 
$f_{compl}$ 
 we will discuss are:

 (A)

$$
 F_{src}(\tau,q,Q) 
  = -\gamma_2 q(\tau)+\gamma_3 \frac{\partial}{\partial \tau}q(\tau)
 \,,\,
 f_{compl}(t,q,Q) = \gamma_1 Q(t) + f_0(t) 
$$

 (B)

$$
 F_{src}(\tau,q,Q) = \gamma_0 (Q(\tau)-q(\tau))
 \,,\,
 f_{compl}(t,q,Q) = \Omega^2 Q(t) + f_0(t)
$$

\newpage\section
{ the Models } 
%
%
\subsection{
 Effective Equation of Motion of the Particle 
 in the Case of (A), i.e., in the Case where 
$$
 F_{src}(\tau,q,Q) 
  = -\gamma_2 q(\tau)+\gamma_3 \frac{\partial}{\partial \tau}q(\tau)
 \,,\,
 f_{compl}(t,q,Q) = \gamma_1 Q(t) + f_0(t) 
$$ 
    }

 In this case the formulae 
\begin{eqnarray*}
 \frac{\partial^2 q(t)}{\partial t^2}
 &=&-\Omega^2 q(t)+f_{compl}(t,q,Q) 
\\
 Q(t)
 &=&
 \Big(\alpha_0 + \alpha_1 \frac{\partial}{\partial t} \Big)
 \Bigg(
 -{2\gamma}
 \int_0^t 
 \Big(F_{src}(\tau,q,Q)\Big)d\tau
 + u_{01}(t,x_0) \Bigg)  
\\
 Q(t)  
 &=&
 \Big(\alpha_0 + \alpha_1 \frac{\partial}{\partial t} \Big)
 \Bigg\{ u(t,x) \Bigg\} \Bigg|_{x=x_0}  
\\
\end{eqnarray*}

$$
\parbox{\textwidth}{
\begin{eqnarray*}
 u(t,x) 
 &=&
 \left\{ 
 \begin{array}{ccl}
 \displaystyle
 -{2\gamma}
 \int_0^{t-|x-x_0|/c} 
 \Big(F_{src}(\tau,q,Q)\Big)d\tau &,& \mbox{ if } 0 \leq t-|x-x_0|/c \\ 
                 0              &,& \mbox{ if } t-|x-x_0|/c < 0 \leq t \\ 
 \end{array}
 \right\} 
 \\[\medskipamount]\\
 &&{}\qquad + u_{01}(t,x) 
 \\[\smallskipamount] \\ 
\end{eqnarray*}
}
$$
 become 

\begin{eqnarray*}
 \frac{\partial^2 q(t)}{\partial t^2}
 &=&-\Omega^2 q(t)+\gamma_1 Q(t) + f_0(t)  
\\
 Q(t)
 &=&
 \Big(\alpha_0 + \alpha_1 \frac{\partial}{\partial t} \Big)
 \Bigg(
 -{2\gamma}
 \int_0^t 
 \Big(-\gamma_2+\gamma_3 \frac{\partial}{\partial \tau}
 \Big) q(\tau)d\tau 
 + u_{01}(t,x_0) \Bigg)  
\\
 Q(t)  
 &=&
 \Big(\alpha_0 + \alpha_1 \frac{\partial}{\partial t} \Big)
 \Bigg\{ u(t,x) \Bigg\} \Bigg|_{x=x_0}  
\\
\end{eqnarray*}

$$
\parbox{\textwidth}{
\begin{eqnarray*}
 u(t,x) 
 &=&
 \left\{ 
 \begin{array}{ccl}
 \displaystyle
 -{2\gamma}
 \int_0^{t-|x-x_0|/c} 
 \Big(-\gamma_2+\gamma_3 \frac{\partial}{\partial \tau}
 \Big) q(\tau)d\tau             &,& \mbox{ if } 0 \leq t-|x-x_0|/c \\ 
                 0              &,& \mbox{ if } t-|x-x_0|/c < 0 \leq t \\ 
 \end{array}
 \right\} 
 \\[\medskipamount]\\
 &&{}\qquad + u_{01}(t,x) 
 \\[\smallskipamount] \\ 
\end{eqnarray*}
}
$$

\newpage 
\noindent 
 As a consequence, 
\begin{eqnarray*}
 \frac{\partial^2 q(t)}{\partial t^2}
 &=&-\Omega^2 q(t)
    +\gamma_1  \Big(\alpha_0 + \alpha_1 \frac{\partial}{\partial t} \Big)
 \Bigg(
 -{2\gamma}
 \int_0^t 
 \Big(-\gamma_2+\gamma_3 \frac{\partial}{\partial \tau}
 \Big) q(\tau)d\tau
 + u_{01}(t,x_0) \Bigg)  
 + f_0(t)  
\\
\end{eqnarray*}
 and then there follow quite regular transformations:
\begin{eqnarray*}
 \frac{\partial^2 q(t)}{\partial t^2}
 &=&-\Omega^2 q(t)
    +\gamma_1  \Big(\alpha_0 + \alpha_1 \frac{\partial}{\partial t} \Big)
 \Bigg(
 -{2\gamma}
 \int_0^t 
 \Big(-\gamma_2+\gamma_3 \frac{\partial}{\partial \tau}
 \Big) q(\tau)d\tau
 \Bigg)  
\\&&{} 
    +\gamma_1  \Big(\alpha_0 + \alpha_1 \frac{\partial}{\partial t} \Big)
 \Bigg(
 u_{01}(t,x_0) \Bigg)  
 + f_0(t)  
\\
\end{eqnarray*}

\begin{eqnarray*}
 \frac{\partial^2 q(t)}{\partial t^2}
 &=&-\Omega^2 q(t)
    +\gamma_1\alpha_0 
 \Bigg(
 -{2\gamma}
 \int_0^t 
 \Big(-\gamma_2+\gamma_3 \frac{\partial}{\partial \tau}
 \Big) q(\tau)d\tau
 \Bigg)  
\\&&{} 
    +\gamma_1\alpha_1 \frac{\partial}{\partial t} 
 \Bigg(
 -{2\gamma}
 \int_0^t 
 \Big(-\gamma_2+\gamma_3 \frac{\partial}{\partial \tau}
 \Big) q(\tau)d\tau
 \Bigg)  
\\&&{} 
    +\gamma_1  \Big(\alpha_0 + \alpha_1 \frac{\partial}{\partial t} \Big)
 \Bigg(
 u_{01}(t,x_0) \Bigg)  
 + f_0(t)  
\\
\end{eqnarray*}
 If 
${\gamma}$
 is a constant in 
$t$
, then 
\begin{eqnarray*}
 \frac{\partial^2 q(t)}{\partial t^2}
 &=&-\Omega^2 q(t)
    -2\gamma_1\alpha_0 
 {\gamma}
 \int_0^t 
 \Big(-\gamma_2+\gamma_3 \frac{\partial}{\partial \tau}
 \Big) q(\tau)d\tau
\\&&{} 
    -2\gamma_1\alpha_1 
 {\gamma}
 \Big(-\gamma_2+\gamma_3 \frac{\partial}{\partial t}
 \Big) q(t)
\\&&{} 
    +\gamma_1  \Big(\alpha_0 + \alpha_1 \frac{\partial}{\partial t} \Big)
 u_{01}(t,x_0)
 + f_0(t)  
\\
\end{eqnarray*}

 We have just obtained an effective equation of motion of the particle 
 subject to the model (A),
 and now,  
 let us now try to solve this equation. 
 We restrict ourselves to the case where 
 the field 
 is initially 
 not excited: 
$$
 u_{01}(t,x_0) = 0.
$$
 So, we now deal with the case where  
\begin{eqnarray*}
 \frac{\partial^2 q(t)}{\partial t^2}
 &=&-\Omega^2 q(t)
    -2\gamma_1\alpha_0 
 {\gamma}
 \int_0^t 
 \Big(-\gamma_2+\gamma_3 \frac{\partial}{\partial \tau}
 \Big) q(\tau)d\tau
\\&&{} 
    -2\gamma_1\alpha_1 
 {\gamma}	
 \Big(-\gamma_2+\gamma_3 \frac{\partial}{\partial t}
 \Big) q(t)
 + f_0(t)  
\\
\end{eqnarray*}
 and we see, in the relation written down, the term that one writes 
 tending to express the idea of linear friction: it is  
$$
 -2\gamma_1\alpha_1{\gamma}\gamma_3 \frac{\partial}{\partial t}
  q(t) \,.
$$
 But another detail attracts attention:
 this relation is not an ORDINARY differential relation 
 whenever 
$$
 \alpha_0 \not= 0 \,.
$$ 
 It is because of the term 
$$
    -2\gamma_1\alpha_0 
 {\gamma}
 \int_0^t 
 \Big(-\gamma_2+\gamma_3 \frac{\partial}{\partial \tau}
 \Big) q(\tau)d\tau
$$
 Only if 
$\alpha_0 = 0$ 
 we see a relation which construction is habitual: 
\begin{eqnarray*}
 \frac{\partial^2 q(t)}{\partial t^2}
 &=&-\Omega^2 q(t)
    -2\gamma_1\alpha_1 
 {\gamma}
 \Big(-\gamma_2+\gamma_3 \frac{\partial}{\partial t}
 \Big) q(t)
 + f_0(t)  
\\
\end{eqnarray*}
 In other cases we deal with the particle's motion 
 that one is used to qualifying 
 as motion ( or, evolution ) with memory.  
 Even if 
$$
 \gamma_2 = 0
$$
 the equation of motion is not ORDINARY differential equation 
 because of 
$q(0)$ 
 in 
$$
-2\gamma_1\alpha_0
 \int_0^t 
 \gamma_3 \frac{\partial}{\partial \tau}
 q(\tau)d\tau
 = 
 -2\gamma_1\alpha_0\gamma_3\Big( q(t) - q(0) \Big)
 \eqno ( \mbox{ if } \gamma_3 = const )
$$ 
 One can say we have models of dynamics with 
 "on only one instant concentrated memory".
 For a contrast, the case where 
$$
 \gamma_2 \not= 0\,,\, \gamma_3 = 0 \,,\,
$$
 can be referred to as a case of 
 "wide memory"

 Let us now consider some particular cases that represent 
 (as we think) 
 the most typical properties of the general case. For simplicity, we assume 
 all 
$\alpha$-s 
 and 
$\gamma$-s 
 to be positive, and constant in 
$t$. 

 \addvspace{\bigskipamount}

 "Habitual" case: 

\begin{eqnarray*}
 \frac{\partial^2 q(t)}{\partial t^2}
 &=&-\Omega^2 q(t)
    -2\gamma_1\alpha_1 
 {\gamma}
 \Big(-\gamma_2+\gamma_3 \frac{\partial}{\partial t}
 \Big) q(t)
 + f_0(t) \,. 
\\
\end{eqnarray*}
 There is no surprise, with the possible exception of the case, where 
$$
 (\gamma_1\alpha_1{\gamma}\gamma_3)^2-\Omega^2 
  +2\gamma_1\alpha_1{\gamma}\gamma_2 
 \geq 0
 \,,\,
$$
$$
 -\gamma_1\alpha_1{\gamma}\gamma_3
 +\sqrt{
 (\gamma_1\alpha_1{\gamma}\gamma_3)^2-\Omega^2 
  +2\gamma_1\alpha_1{\gamma}\gamma_2} 
 \geq 0
$$
 In such a case we observe self-acceleration of the oscillator, 
 a phenomenon discussed and been discussing at least in electrodynamics. 

 \addvspace{\bigskipamount}

 The second case is 
 the case of  
 "on only one instant concentrated memory":

\begin{eqnarray*}
 \frac{\partial^2 q(t)}{\partial t^2}
 &=&-\Omega^2 q(t)
    -2\gamma_1\alpha_0 
 {\gamma}
 \gamma_3 
 \Big(q(t)-q(0)\Big)
\\&&{} 
    -2\gamma_1\alpha_1 
 {\gamma}
 \gamma_3 \frac{\partial}{\partial t}
 q(t)
 + f_0(t)  
\\
\end{eqnarray*}

 As we have recently pointed up, 
 this equation is not ordinary differential one.
 However, the machinery of the {\bf ordinary} differential equations 
 does here {\bf quite for}.
 We illustrate it by the example, where  
$$
 f_0(t) =0 \,,
$$
 i.e., where 
\begin{eqnarray*}
 \frac{\partial^2 q(t)}{\partial t^2}
 &=&-\Omega^2 q(t)
    -2\gamma_1\alpha_0 
 {\gamma}
 \gamma_3 
 \Big(q(t)-q(0)\Big)
\\&&{} 
    -2\gamma_1\alpha_1 
 {\gamma}
 \gamma_3 \frac{\partial}{\partial t}
 q(t)
\\
\end{eqnarray*}
 We emphasise: this equation is linear {\em homogeneous } in 
$q(t)$, 
 but write it as 
\begin{eqnarray*}
 \frac{\partial^2 q(t)}{\partial t^2}
 +
 2\gamma_1\alpha_1
 {\gamma}
 \gamma_3 \frac{\partial q(t)}{\partial t}
 +(\Omega^2 +2\gamma_1\alpha_0{\gamma}\gamma_3 ) q(t)
 &=&
 2\gamma_1\alpha_0{\gamma}\gamma_3 q(0)
\\
\end{eqnarray*}
 and consider this equation as {\em inhomogeneous} one. 

 Thus, we infer that
$$
 q(t) 
 =\frac{2\gamma_1\alpha_0{\gamma}\gamma_3}
  {\Omega^2 +2\gamma_1\alpha_0{\gamma}\gamma_3} 
 q(0)
 +e^{-\gamma_1\alpha_1{\gamma}\gamma_3 t}
 (C_c\cos(\Omega_g t) + C_s\sin(\Omega_g t))
$$
 where 
$$
 \Omega_g 
 :=
 \sqrt{
 \Omega^2 +2\gamma_1\alpha_0{\gamma}\gamma_3
 -(\gamma_1\alpha_1{\gamma}\gamma_3)^2}
$$
 At first, we can calculate 
$C_c$. 
 Actually, we have  
$$
 q(0) 
 =\frac{2\gamma_1\alpha_0{\gamma}\gamma_3}
  {\Omega^2 +2\gamma_1\alpha_0{\gamma}\gamma_3} 
 q(0)
 + C_c \, 
$$
 i.e., 
$$
 C_c = q(0)-\frac{2\gamma_1\alpha_0{\gamma}\gamma_3}
  {\Omega^2 +2\gamma_1\alpha_0{\gamma}\gamma_3} q(0) 
$$
 Thus we have calculated 
$C_c$.
 As for 
$C_s$, 
 it can be similarly calculated. 

 An interesting detail is: 
 if 
$t\to +\infty$ 
 then 
$q(t)$ 
 has a limit, 
 in this sense 
$q(t)$ 
 behaves as the usual damped oscillator. But 
$$
 q(t)
 \to 
 \frac{2\gamma_1\alpha_0{\gamma}\gamma_3}
  {\Omega^2 +2\gamma_1\alpha_0{\gamma}\gamma_3} 
 q(0)
 \not= 0 \qquad 
 \mbox{ every time that 
$q(0) \not= 0$
  and 
$\gamma_1\alpha_0{\gamma}\gamma_3 \not= 0$
 {\large \bf !!!}}
$$ 
 We observe an element of a plastic behaviour!

 \addvspace{\bigskipamount}

 The last particular case we wish to discuss is 
 the case of 
 "wide memory":

\begin{eqnarray*}
 \frac{\partial^2 q(t)}{\partial t^2}
 &=&-\Omega^2 q(t)
    +2\gamma_1\alpha_0 
 {\gamma}
 \gamma_2
 \int_0^t 
 q(\tau)d\tau
\\&&{} 
    +2\gamma_1\alpha_1 
 {\gamma}
 \gamma_2 
 q(t)
 + f_0(t)  
\\
\end{eqnarray*}
 Technically, this is the most complicated case, 
 with the possible exception of the general one, for that reason we 
 restrict ourselves to case where  
$$
 f_0(t)=0\,,\, \alpha_1=0 \,,
$$
 i.e., where 
\begin{eqnarray*}
 \frac{\partial^2 q(t)}{\partial t^2}
 &=&-\Omega^2 q(t)
    +2\gamma_1\alpha_0{\gamma}\gamma_2
 \int_0^t 
 q(\tau)d\tau
\\
\end{eqnarray*}
 and concentrate only on the system's behaviour at large  
$t$ . 
 In order to estimate the asymptotic behaviour of 
$q(t)$ 
 as 
$ t \to +\infty $,
 let us handle with the characteristic polynomials. It is:

$$
 \lambda^2 +\Omega^2 - 2\gamma_1\alpha_0{\gamma}{\gamma_2}\frac{1}{\lambda} 
 =
 (\lambda^3 +\Omega^2\lambda - 2\gamma_1\alpha_0{\gamma}{\gamma_2})/\lambda 
$$ 
 The situation is dramatic. The polynomial 
$$
 \lambda^3 +\Omega^2\lambda - 2\gamma_1\alpha_0{\gamma}{\gamma_2} 
$$ 
 has one pure real root 
$\lambda_{1}$ 
 and two complex-conjugated ones: 
$\lambda_{2}\,,\lambda_{3} \,,\,\lambda_{3} = \overline{\lambda_{2}}$ .
 Since 
$$
  \lambda_{1} +\lambda_{2} +\lambda_{3}  = 0 
 \,,\,
  \lambda_{1} \lambda_{2} \lambda_{3}  = 2\gamma_1\alpha_0{\gamma}{\gamma_2} 
$$ 
 we have 
$$
  \lambda_{1} +2Re\lambda_{2} = 0 
 \,,\,
  \lambda_{1} |\lambda_{2}|^2  = 2\gamma_1\alpha_0{\gamma}{\gamma_2} 
$$ 
 and hence 
$$
 \lambda_{1} > 0\,\mbox{ (!!!) },
 Re\lambda_{2} = Re\lambda_{3} <0 \,. 
$$
 Thus, we expect an EXPONENTIAL GROWTH of the oscillator amplitude, 
 as 
$t\to +\infty$. 
 Does this exponential growth really exist? 
 Actually, 
 any expression 
$$
  C_1 e^{\lambda_1 t} +C_2 e^{\lambda_2 t} +C_3 e^{\overline{\lambda_2} t}
 \,,
$$
 as a function of 
$t$, 
 satisfies the relation
\begin{eqnarray*}
 \frac{\partial^3 q(t)}{\partial t^3}
 &=&-\Omega^2 \frac{\partial q(t)}{\partial t}
    +2\gamma_1\alpha_0{\gamma}\gamma_2
 q(t) \,.
\\
\end{eqnarray*}
 But it in itself does not mean that 
 such an expression satisfies the proper relation
\begin{eqnarray*}
 \frac{\partial^2 q(t)}{\partial t^2}
 &=&-\Omega^2 q(t)
    +2\gamma_1\alpha_0{\gamma}\gamma_2
 \int_0^t 
 q(\tau)d\tau  \,.
\\
\end{eqnarray*}
 A priori, all that we can now assert is that if a function, 
$ F $
 of 
$t$
 , is of the form 
$$
 F(t) 
 = C_1 e^{\lambda_1 t} +C_2 e^{\lambda_2 t} +C_3 e^{\overline{\lambda_2} t}
 \,,
$$
 then 
\begin{eqnarray*}
 \frac{\partial^2 F(t)}{\partial t^2}
 &=&-\Omega^2 F(t)
    +2\gamma_1\alpha_0{\gamma}\gamma_2
 \int_0^t 
 F(\tau)d\tau + \makebox{{\it a Constant }}\,.
\\
\end{eqnarray*}

 Fortunately, the problem, that has just arisen, is not difficult.
 Of course, every proper 
$q(t)$ 
 is of the form 
$$
  C_1 e^{\lambda_1 t} +C_2 e^{\lambda_2 t} +C_3 e^{\overline{\lambda_2} t}
 \,.
$$
 Moreover, 
$$
 q(t) 
 = C_1 e^{\lambda_1 t} +C_2 e^{\lambda_2 t} 
   +\overline{C_2} e^{\overline{\lambda_2} t}
$$
 because 
$q(t)$ 
 is a particle position, hence, 
$q(t)$ 
 is real.
 Thus, if we take into account that 
\begin{eqnarray*}
\makebox[10ex][l]{$\displaystyle 
 \int_0^t 
  C_1 e^{\lambda_1 \tau} +C_2 e^{\lambda_2 \tau} 
  +C_2 e^{\overline{\lambda_2} \tau}
  d\tau
$}
 \\&=&
 \frac{C_1}{\lambda_1} e^{\lambda_1 t} 
 +\frac{C_2}{\lambda_2} e^{\lambda_2 t} 
  +\overline{\Bigl(\frac{C_2}{\lambda_3}\Bigr)} e^{\overline{\lambda_2} t}
 -
 \Bigl(
 \frac{C_1}{\lambda_1}
 +\frac{C_2}{\lambda_2}
  +\overline{\Bigl(\frac{C_2}{\lambda_3}\Bigr)}
\Bigr)
 \\&=&
 \frac{C_1}{\lambda_1} e^{\lambda_1 t} 
 +\frac{C_2}{\lambda_2} e^{\lambda_2 t} 
  +\overline{\Bigl(\frac{C_2}{\lambda_3}\Bigr)} e^{\overline{\lambda_2} t}
 -
 \Bigl(
 \frac{C_1}{\lambda_1}
 +2Re\Bigl(\frac{C_2}{\lambda_2}\Bigr)
 \Bigr)
\\
\end{eqnarray*}
 we can conclude: if 
$$
  \frac{C_1}{\lambda_1}
 +2Re\Bigl(\frac{C_2}{\lambda_2}\Bigr) = 0 
$$
 then 
\begin{eqnarray*}
 q(t) 
 &:=&
  C_1 e^{\lambda_1 t} +C_2 e^{\lambda_2 t} 
   +\overline{C_2} e^{\overline{\lambda_2} t} \,,
\\&\equiv &
  -2Re\Bigl(\frac{C_2}{\lambda_2}\Bigr) e^{\lambda_1 t} +C_2 e^{\lambda_2 t} 
   +\overline{C_2} e^{\overline{\lambda_2} t}
\end{eqnarray*}
 satisfies the proper relation. 
 Thus, if we take 
$C_2$
 so that
$$
 2Re\Bigl(\frac{C_2}{\lambda_2}\Bigr) \not= 0 \,,
$$
 e.g.,
$C_2 = \lambda_2$,
 then 
$$
 q(t) \asymp -2Re\Bigl(\frac{C_2}{\lambda_2}\Bigr) e^{\lambda_1 t} 
 \makebox{ as }  
 t \to +\infty \,.
$$

 Thus, we really observe an EXPONENTIAL GROWTH of the oscillator amplitude, 
 as 
$t\to +\infty$, 
 a factor that can throw Physicist's mind into confusion. 
 We cannot here hope we have simply confounded the `time directions'.
 If we had, 
 we would have {\bf two} roots with strictly positive real parts! 
 In any case, we cannot escape from phenomenon of "self-acceleration". 
 We defer the more detailed discussion on this subject
 and notice only, that
 a related phenomenon is known in electrodynimics, see 
 Abraham-Lorentz-Dirac equations.

\addvspace{\bigskipamount}


\addvspace{\bigskipamount}

 \hspace*{\fill} 
 **********************************************************
 \hspace*{\fill} 

\addvspace{\bigskipamount}

 Now then, 
 we have obtained an effective equation of motion of the particle 
 subject to the model (A) 
 and discussed properties of this model, 
 and we turn now to the model (B) with the same intention.

\newpage 
\subsection{ 
 Effective Equation of Motion of the Particle 
 in the Case of (B), i.e., in the Case where 
$$
 F_{src}(\tau,q,Q) = \gamma_0 (Q(\tau)-q(\tau))
 \,,\,
 f_{compl}(t,q,Q) = \Omega^2 Q(t) + f_0(t)
$$ 
}

 In this case the formulae 
\begin{eqnarray*}
 \frac{\partial^2 q(t)}{\partial t^2}
 &=&-\Omega^2 q(t)+f_{compl}(t,q,Q) 
\\
 Q(t)
 &=&
 \Big(\alpha_0 + \alpha_1 \frac{\partial}{\partial t} \Big)
 \Bigg(
 -{2\gamma}
 \int_0^t 
 \Big(F_{src}(\tau,q,Q)\Big)d\tau
 + u_{01}(t,x_0) \Bigg)  
\\
 Q(t)  
 &=&
 \Big(\alpha_0 + \alpha_1 \frac{\partial}{\partial t} \Big)
 \Bigg\{ u(t,x) \Bigg\} \Bigg|_{x=x_0}  
\\
\end{eqnarray*}
$$
\parbox{\textwidth}{
\begin{eqnarray*}
 u(t,x) 
 &=&
 \left\{ 
 \begin{array}{ccl}
 \displaystyle
 -{2\gamma}
 \int_0^{t-|x-x_0|/c} 
 \Big(F_{src}(\tau,q,Q)\Big)d\tau &,& \mbox{ if } 0 \leq t-|x-x_0|/c \\ 
                 0              &,& \mbox{ if } t-|x-x_0|/c < 0 \leq t \\ 
 \end{array}
 \right\} 
 \\[\medskipamount]\\
 &&{}\qquad + u_{01}(t,x) 
 \\[\smallskipamount] \\ 
\end{eqnarray*}
}
$$
 become 

\begin{eqnarray*}
 \frac{\partial^2 q(t)}{\partial t^2}
 &=&-\Omega^2 q(t)+\Omega^2 Q(t) + f_0(t) 
\\
 Q(t)
 &=&
 \Big(\alpha_0 + \alpha_1 \frac{\partial}{\partial t} \Big)
 \Bigg(
 -{2\gamma}
 \int_0^t 
 \Big(\gamma_0 (Q(\tau)-q(\tau))\Big)d\tau
 + u_{01}(t,x_0) \Bigg)  
\\
 Q(t)  
 &=&
 \Big(\alpha_0 + \alpha_1 \frac{\partial}{\partial t} \Big)
 \Bigg\{ u(t,x) \Bigg\} \Bigg|_{x=x_0}  
\\
\end{eqnarray*}

$$
\parbox{\textwidth}{
\begin{eqnarray*}
 u(t,x) 
 &=&
 \left\{ 
 \begin{array}{ccl}
 \displaystyle
 -{2\gamma}
 \int_0^{t-|x-x_0|/c} 
 \Big(\gamma_0 (Q(\tau)-q(\tau))\Big)d\tau 
                                &,& \mbox{ if } 0 \leq t-|x-x_0|/c \\ 
                 0              &,& \mbox{ if } t-|x-x_0|/c < 0 \leq t \\ 
 \end{array}
 \right\} 
 \\[\medskipamount]\\
 &&{}\qquad + u_{01}(t,x) 
 \\[\smallskipamount] \\ 
\end{eqnarray*}
}
$$

\newpage 
\noindent
 Let us focus our attention 
 firstly on 
\begin{eqnarray*}
 \frac{\partial^2 q(t)}{\partial t^2}
 &=&-\Omega^2 q(t)+\Omega^2 Q(\tau) + f_0(t)
\\
 Q(t)
 &=&
 \Big(\alpha_0 + \alpha_1 \frac{\partial}{\partial t} \Big)
 \Bigg(
 -{2\gamma}
 \int_0^t 
 \Big(\gamma_0 (Q(\tau)-q(\tau))\Big)d\tau
 + u_{01}(t,x_0) \Bigg)  
\\
\end{eqnarray*}

  Write it as 

\begin{eqnarray*}
 \frac{\partial^2 q(t)}{\partial t^2} - f_0(t)
 &=& -\Omega^2 (q(t)-Q(\tau)) 
\\
 Q(t)
 +\Big(\alpha_0 + \alpha_1 \frac{\partial}{\partial t} \Big)
 {2\gamma}
 \int_0^t 
 \Big(\gamma_0 Q(\tau)\Big)d\tau
 &=&
 \Big(\alpha_0 + \alpha_1 \frac{\partial}{\partial t} \Big)
 \Bigg(
 {2\gamma}
 \int_0^t 
 \Big(\gamma_0 q(\tau)\Big)d\tau
 + u_{01}(t,x_0) \Bigg)  
\\
\end{eqnarray*}

 We have now obtained an equation array for 
$Q$ and $q$. 
 The next step is to form an {\bf insulated} equation for 
$q$. 
\footnote{ as for an {\bf insulated} equation for 
$Q$,
 one can find it in Appendix A.
}
 For this purpose, we apply, at first, the operator 
$$
  \Big( I + 
  \Big(\alpha_0 + \alpha_1 \frac{\partial}{\partial t} \Big)
  {2\gamma} 
  \int_0^t\gamma_0 \cdot d\tau  \Big)
$$
 to the former equation, i.e., to the equation 
\begin{eqnarray*} 
  \frac{\partial^2 q}{\partial t^2} - f_0
 &=& -\Omega^2(q-Q) 
\end{eqnarray*}
 Then we infer 

\begin{eqnarray*}
\makebox[26ex][l]{$\displaystyle 
  \Big(
  \frac{\partial^2 q(t)}{\partial t^2} - f_0(t)
  \Big)
 +
 \Big(\alpha_0 + \alpha_1 \frac{\partial}{\partial t} \Big)
  {2\gamma} \int_0^t 
  \gamma_0 
  \Big( 
  \frac{\partial^2 q(\tau)}{\partial \tau^2} - f_0(t) 
  \Big)d\tau 
$}
\\&=&
 -\Omega^2 \Big( q(t) 
  +
 \Big(\alpha_0 + \alpha_1 \frac{\partial}{\partial t} \Big)
  {2\gamma}
  \int_0^t \gamma_0 q(\tau) d\tau \Big)
\\&&{} 
 +\Omega^2 \Big( Q(t) 
  +
 \Big(\alpha_0 + \alpha_1 \frac{\partial}{\partial t} \Big)
  {2\gamma} \int_0^t \gamma_0 Q(\tau) d\tau \Big)
\\[\bigskipamount]\\ 
\makebox[26ex][l]{$\displaystyle 
  \Big(
  \frac{\partial^2 q(t)}{\partial t^2} - f_0(t)
  \Big)
 + 
 \Big(\alpha_0 + \alpha_1 \frac{\partial}{\partial t} \Big)
  {2\gamma} \int_0^t 
  \gamma_0 
  \Big(
  \frac{\partial^2 q(\tau)}{\partial \tau^2} - f_0(t) 
  \Big)d\tau 
$}
\\&=&
 -\Omega^2 \Big( q(t) 
  +
 \Big(\alpha_0 + \alpha_1 \frac{\partial}{\partial t} \Big)
 {2\gamma} \int_0^t \gamma_0 q(\tau) d\tau \Big)
\\&&{}
 +\Omega^2 \Big(
 \Big(\alpha_0 + \alpha_1 \frac{\partial}{\partial t} \Big)
 {2\gamma} \int_0^t \gamma_0 q(\tau) d\tau
 + u_{01}(t,x_0) 
 \Big)
\\
\end{eqnarray*}

\addvspace{\medskipamount} 

$$
\fbox{
\parbox{\textwidth}{
\begin{eqnarray*}
\makebox[50ex][l]{$\displaystyle 
  \Big(
  \frac{\partial^2 q(t)}{\partial t^2} - f_0(t)
  \Big)
  +
 \Big(\alpha_0 + \alpha_1 \frac{\partial}{\partial t} \Big)
  {2\gamma}
  \int_0^t 
  \gamma_0
  \Big(
  \frac{\partial^2 q(\tau)}{\partial \tau^2} - f_0(t) 
  \Big)d\tau 
 $}
\\&=&
 -\Omega^2
 \Big( q(t) + u_{01}(t,x_0) \Big)
\\
\end{eqnarray*}
}
}
$$

\addvspace{\medskipamount} 

 or, 

\addvspace{\medskipamount} 

$$
\fbox{
\parbox{0.75\textwidth}{
\begin{eqnarray*}
 \Big(
 (I + K)(\ddot q - f_0)
 \Big)
 (t)
 &=&
 -\Omega^2
 \Big( q(t) + u_{01}(t,x_0) \Big)
\\
\end{eqnarray*}
 where 
$$
 K 
  := 
 \Big(\alpha_0 + \alpha_1 \frac{\partial}{\partial t} \Big)
 {2\gamma} 
 \int_0^t\gamma_0 \cdot d\tau  
 \,,\qquad 
 \ddot q(t) := \frac{\partial^2 q(t)}{\partial t^2}
$$
}
}
\footnote{ general abstractions are evident!}
$$

\addvspace{2\bigskipamount} 

 We can stop at this last equation, or, observing that 
$$ 
 \int_0^t  \frac{\partial^2 q(\tau)}{\partial \tau^2}d\tau  
 = \frac{\partial q(t)}{\partial t}
   -\frac{\partial q(t)}{\partial t}\Big|_{t=0}
 \,, 
$$
 we can stop at that:

\medskip 

$$
\fbox{
\parbox{\textwidth}{
\begin{eqnarray*}
\makebox[20ex][l]{$\displaystyle 
 (1+2\gamma\gamma_0\alpha_1)
 \Big(
 \frac{\partial^2 q(t)}{\partial t^2} - f_0(t)
 \Big)
 $}
\\&=&
   -{2\gamma\gamma_0}\alpha_0\frac{\partial q(t)}{\partial t} 
   -\Omega^2 q(t) 
   +{2\gamma\gamma_0}\alpha_0\frac{\partial q(t)}{\partial t}\Big|_{t=0} 
\\&&{}
   + \Omega^2 u_{01}(t,x_0) 
   +{2\gamma\gamma_0}\alpha_0 \int_0^t f_0(\tau) d\tau 
\end{eqnarray*}
}
}
$$
\medskip 

\noindent 
  Some people prefer to write such an equation as following: 

\medskip 

$$
\fbox{
\parbox{\textwidth}{
\begin{eqnarray*}
\makebox[2ex][l]{$\displaystyle 
 \Bigl(
 (1+2\gamma\gamma_0\alpha_1)
 \frac{\partial^2 }{\partial t^2} 
  +{2\gamma\gamma_0}\alpha_0\frac{\partial }{\partial t} 
  +\Omega^2 \Bigr)q(t) 
 $}
\\&=&
   {2\gamma\gamma_0}\alpha_0
   \frac{\partial q(t)}{\partial t}\Big|_{t=0} 
   + \Omega^2 u_{01}(t,x_0) 
   + (1+2\gamma\gamma_0\alpha_1)f_0(t) 
   +{2\gamma\gamma_0}\alpha_0 \int_0^t f_0(\tau) d\tau 
\end{eqnarray*}
}
}
$$

\bigskip 

 We have just obtained an effective equation of motion of the particle 
 subject to the model (B),
 and now, as in the previous subsection, 
 we are trying to solve the resulting equation. 
 As in the previous subsection, 
 we restrict ourselves to the case where 
 the field 
 is initially 
 in unexcited state: 
$$
 u_{01}(t,x_0) = 0.
$$
 In addition, for simplicity, we take 
$$
 \gamma_0 :=1\,,\,\alpha_0 :=1\,,\,\alpha_1 :=0\,.
$$
 So, we now deal with the case where  
$$
 \Bigl(\frac{\partial^2 }{\partial t^2} 
  +{2\gamma}\frac{\partial }{\partial t} 
  +\Omega^2 \Bigr)q(t) 
  =
   {2\gamma}\frac{\partial q(t)}{\partial t}\Big|_{t=0} 
   + f_0(t) 
   +{2\gamma} \int_0^t f_0(\tau) d\tau 
$$ 

 We start out emphasising that the homogeneous equation, connected
 to this equation 
 is exactly  
\begin{eqnarray*} 
 \Bigl(\frac{\partial^2 }{\partial t^2} 
  +{2\gamma}\frac{\partial }{\partial t} 
  +\Omega^2 \Bigr)q(t) 
  &=&
   {2\gamma}\frac{\partial q(t)}{\partial t}\Big|_{t=0} 
 \\
\end{eqnarray*}
 and NOT 
\begin{eqnarray*} 
 \Bigl(\frac{\partial^2 }{\partial t^2} 
  +{2\gamma}\frac{\partial }{\partial t} 
  +\Omega^2 \Bigr)q(t) 
 &=& 
 0 
 \\ 
\end{eqnarray*}
 In the previous subsection 
\footnote{ and in subsection 1.1, as well!}, 
 we have already discussed the similar factor, 
 and we are using the similar machinery. 
 The difference 
 between the equations 
 we have just now written 
 is the {\bf rank one } term 
${2\gamma}\frac{\partial q(t)}{\partial t}\Big|_{t=0}$ 
\footnote{ 
 This term, as a function of 
$t$ ,
 is a FIXED function of 
$t$, 
 in this context a constant non-zero function, 
 e.g. 
$1$, multiplied by a CONSTANT depended on $q$, i.e.,
 by a fixed functional of 
$q$. 
 Using the Dirac's syntax, one can write this term as 
$|a><b|$ 
 with 
$|a>=1$
 and
$<b|q>={2\gamma}\frac{\partial q(t)}{\partial t}\Big|_{t=0}$ . 
 }
 . 
 This detail allows us to apply 
 the usual machinery of the finite rank perturbations theory. 
 Thus, having put 
$$
 \Omega_{\gamma} := \sqrt{\Omega^2 -{\gamma}^2}
$$
 and having taken into account the reasons of the previous subsections, 
 we can show that
\begin{eqnarray*}
 q(t) 
 &=& 
   e^{-\gamma t} 
  \Big(cos(\Omega_{\gamma} t) 
        +\gamma \frac{sin(\Omega_{\gamma} t)}{\Omega_{\gamma}}
  \Big)
  \Big(q(0) 
      -\frac{2\gamma}{\Omega^2}\frac{\partial q(t)}{\partial t}\Big|_{t=0}  
  \Big)
   + e^{-\gamma t}\frac{sin(\Omega_{\gamma} t)}{\Omega_{\gamma}} 
            \frac{\partial q(t)}{\partial t}\Big|_{t=0}  
 \\&&{ }
   +\frac{2\gamma}{\Omega^2}\frac{\partial q(t)}{\partial t}\Big|_{t=0}  
 \,,
\end{eqnarray*}
 of course, in the case where 
$$
 f_0(t) = 0 \,.
$$ 

\addvspace{2\bigskipamount}

  If we now concentrate on the system's behaviour at large  
$t$ , 
 and where again, for simplicity, 
$f_0(t) = 0$, 
 a mathematical detail calls attention. 
 We observe:
 $$
 q(t)\to  {2\gamma}\frac{\partial q(t)}{\partial t}\Big|_{t=0} 
 \mbox{ as } t \to +\infty \,. 
$$  
 Again, 
 as in the previous subsection, we observe an element of plastic behaviour. 
 A surprising detail in the new situation is: 
 the limit function 
$$
 q_{\infty}(t)={2\gamma}\frac{\partial q(t)}{\partial t}\Big|_{t=0} 
$$
 is NO solution to 
$$
 \frac{\partial^2 q(t)}{\partial t^2} 
  =-{2\gamma}\frac{\partial q(t)}{\partial t} 
   -\Omega^2 q(t) 
   +{2\gamma}\frac{\partial q(t)}{\partial t}\Big|_{t=0} 
$$ 
 every time that 
$$
  {2\gamma}\frac{\partial q(t)}{\partial t}\Big|_{t=0} \not= 0\,, 
$$
 because 
 the 
$q=q_{\infty}$ 
 is a constant, hence its derivative is zero: 
$$
 \frac{\partial q(t)}{\partial t}\Big|_{t=0} 
 = \frac{\partial q_{\infty}}{\partial t} 
 = \frac{\partial }{\partial t} 
   \Bigl( 
   {2\gamma}\frac{\partial q(t)}{\partial t}\Big|_{t=0} 
   \Bigr) = 0 \,. 
$$
 Similar phenomena, 
 one can detect them in the electrodynamics of moving charges.

\addvspace{\bigskipamount}

 \hspace*{\fill} 
 *********************************************************
 \hspace*{\fill} 

\addvspace{\bigskipamount}

 Now then, 
 we have obtained effective equations of motion of the particles 
 subject to the model (A) and, resp., to the model (B), 
 and discussed properties of these models.  

 We have seen that the linear friction can actually be described 
 as a result of radiation reaction. In addition we have seen
 a very simple model of plastic behaviour of dynamical system 
 (section 2.1) 
 and a little more complicated model displaying the same 
 effects (section 2.2). But the specific properties of these two 
 models are different. 
 In the first model memory is a function of initial position, 
 whereas 
 in the second model we rather deal with a function of initial velocity. 
 The memory effects in the two models have an interesting specificity: 
 the moving particle `keeps in its memory' only initial data and `forgets'
 the rest ones, with the possible exception of the `past immediate': 
 really, 
 one needs this `past' to calculate the derivatives of the position, i.e.,
 velocity and acceleration! 
 \ldots 
 We have also seen a model of dynamical system 
 with `wide memory' and self-acceleration \ldots 
  
 Of course, 
 there are other interesting properties of the models, which are presented,
 and many interesting abstractions and generalizations are possible.  
 Nevertheless, 
 it does not form the subject of this paper.

\addvspace{\bigskipamount}

 \hspace*{\fill} 
 *********************************************************
 \hspace*{\fill} 

\addvspace{\bigskipamount}

\newpage 
\section{Appendix A. An insulated equation to $Q(t)$ }  
 After  
$q(t)$
 is found,
 we can determine 
$Q(t)$ ,
 at least formally, by solving 
\begin{eqnarray*}
 Q(t)
 &=&
 \Big(\alpha_0 + \alpha_1 \frac{\partial}{\partial t} \Big)
 \Bigg(
 -{2\gamma}
 \int_0^t 
 \Big(\gamma_0 (Q(\tau)-q(\tau))\Big)d\tau
 + u_{01}(t,x_0) \Bigg)  
 \\
\end{eqnarray*}
 or 
\begin{eqnarray*}
 \frac{\partial Q(t)}{\partial t}
 &=&
 \Big(\alpha_0 + \alpha_1 \frac{\partial}{\partial t} \Big)
 \Bigg(
 -{2\gamma}
 \Big(\gamma_0 (Q(t)-q(t))\Big)
 + u_{01}(t,x_0) \Bigg)  
 \,,
\\&&
 \quad Q(0) 
 =
  \Big(\alpha_0 + \alpha_1 \frac{\partial}{\partial t} \Big)
 u_{01}(t,x_0)\Bigg|_{t=0}
 + \alpha_1 
 \Bigg(
 -{2\gamma}
 \Big(\gamma_0 (Q(0)-q(0))\Big)
 \Bigg)
 \\
\end{eqnarray*}
 
 Thus we have already reduced our model, a model of an oscillator coupled the 
 a scalar field, to a pair of linear `ordinary' differential equations. 
 Nevertheless we want to continue to analyse the matter 
 and we now go searching for another relationships, 
 which would simplify calculations of 
$q, Q$
 and 
$u$.


%
 At first, we will obtain another insulated equation for 
$Q$,
 differently and in a different form. 

 We have 
\begin{eqnarray*}
 \frac{\partial^2 q(t)}{\partial t^2} - f_0(t)
 &=& -\Omega^2 (q(t)-Q(\tau)) 
\\
\makebox[30ex][l]{$\displaystyle 
 Q(t)
 +\Big(\alpha_0 + \alpha_1 \frac{\partial}{\partial t} \Big)
 {2\gamma}
 \int_0^t 
 \Big(\gamma_0 Q(\tau)\Big)d\tau
 $}
\\&=&
 \Big(\alpha_0 + \alpha_1 \frac{\partial}{\partial t} \Big)
 \Bigg(
 {2\gamma}
 \int_0^t 
 \Big(\gamma_0 q(\tau)\Big)d\tau
 + u_{01}(t,x_0) \Bigg) \,, 
\\
\end{eqnarray*}
 i.e.,

\begin{eqnarray*}
 \frac{\partial^2 q(t)}{\partial t^2} - f_0(t)
 &=& -\Omega^2 (q(t)-Q(\tau)) 
\\
\makebox[30ex][l]{$\displaystyle 
 Q(t)-u_{01}(t,x_0)
 +\Big(\alpha_0 + \alpha_1 \frac{\partial}{\partial t} \Big)
 {2\gamma}
 \int_0^t 
 \gamma_0 
 \Big(Q(\tau)-u_{01}(\tau,x_0)\Big)d\tau
 $}
\\&=&
 \Big(\alpha_0 + \alpha_1 \frac{\partial}{\partial t} \Big)
 \Bigg(
 {2\gamma}
 \int_0^t 
 \gamma_0 
 \Big(q(\tau)-u_{01}(\tau,x_0)\Big)d\tau
 \Bigg)  
\\
\end{eqnarray*}
 Denote now, to be more concise, 
\begin{eqnarray*}
 Q_d(t) := Q(t)-u_{01}(t,x_0)
 &,&
 D_{01} := \Big(\alpha_0 + \alpha_1 \frac{\partial}{\partial t} \Big)
 \\
\end{eqnarray*}
 Then we infer 
\begin{eqnarray*} 
\makebox[20ex][l]{$\displaystyle 
 \Big(\frac{\partial^2 }{\partial t^2} + \Omega^2 \Big) 
 \Big(Q_d(t)
  + 
 D_{01}
 {2\gamma} \int_0^t \gamma_0 Q_d(\tau) d\tau \Big) 
 $}
\\&=&
 \Big(\frac{\partial^2 }{\partial t^2} + \Omega^2 \Big)
 D_{01}
 \Bigg({2\gamma} \int_0^t 
 \gamma_0 
 \Big(q(\tau)-u_{01}(\tau,x_0)\Big) d\tau 
 \Bigg)
 \\
\end{eqnarray*}
 and 
\begin{eqnarray*} 
\makebox[5ex][l]{$\displaystyle 
 \Big(\frac{\partial^2 }{\partial t^2} + \Omega^2 \Big) 
 \Big(Q_d(t)
  + D_{01}{2\gamma} \int_0^t \gamma_0 Q_d(\tau) d\tau \Big) 
 $}
\\&=&
 \Big(\frac{\partial^2 }{\partial t^2} + \Omega^2 \Big)
 \Big(D_{01}{2\gamma} \int_0^t \gamma_0 q(\tau) d\tau \Big)
 - 
 \Big(\frac{\partial^2 }{\partial t^2} + \Omega^2 \Big)
 \Big(D_{01}{2\gamma} \int_0^t \gamma_0 u_{01}(\tau,x_0) d\tau \Big)
 \\
\end{eqnarray*}
 Note 
$$
 \frac{\partial^2 }{\partial t^2} \int_0^t \gamma_0 q(\tau) d\tau 
 =
 \frac{\partial }{\partial t} \gamma_0 q(t)  
 =
 \int_0^t\frac{\partial^2 }{\partial \tau^2} \gamma_0 q(\tau) d\tau 
 +
 \frac{\partial }{\partial t} \gamma_0 q(t) \Big|_{t=0} 
 \,,\, 
$$
$$
 \Big(\frac{\partial^2 }{\partial t^2} + \Omega^2 \Big) 
 D_{01}
  =
 D_{01}
 \Big(\frac{\partial^2 }{\partial t^2} + \Omega^2 \Big) 
$$
 Then 
\begin{eqnarray*} 
\makebox[0ex][l]{$\displaystyle 
 \Big(\frac{\partial^2 }{\partial t^2} + \Omega^2 \Big) 
 \Big(Q_d(t)
  + {2\gamma} \int_0^t \gamma_0 Q_d(\tau) d\tau \Big) 
  $}
\\&=&
 D_{01}
 \Bigg(
 {2\gamma} \int_0^t  
 \Big(\frac{\partial^2 }{\partial \tau^2} + \Omega^2 \Big)
 \gamma_0 q(\tau) d\tau 
 +
 {2\gamma}\frac{\partial }{\partial t} \gamma_0 q(t) \Big|_{t=0} 
 - 
 \Big(\frac{\partial^2 }{\partial t^2} + \Omega^2 \Big)
 \Big({2\gamma} \int_0^t \gamma_0 u_{01}(\tau,x_0) d\tau \Big)
 \Bigg)
 \\
\end{eqnarray*}
 Then 
\begin{eqnarray*} 
\makebox[0ex][l]{$\displaystyle 
 \Big(\frac{\partial^2 }{\partial t^2} + \Omega^2 \Big) 
 \Big(Q_d(t)
  + D_{01}{2\gamma} \int_0^t \gamma_0 Q_d(\tau) d\tau \Big) 
  $}
\\&=&
 D_{01}
 \Bigg(
 {2\gamma} \int_0^t  
 \gamma_0 
 \Big( \Omega^2 Q(\tau)  + f_0(\tau) \Big) d\tau 
 +
 {2\gamma}\frac{\partial }{\partial t} \gamma_0 q(t) \Big|_{t=0} 
 - 
 \Big(\frac{\partial^2 }{\partial t^2} + \Omega^2 \Big)
 \Big({2\gamma} \int_0^t \gamma_0 u_{01}(\tau,x_0) d\tau \Big)
 \Bigg)
 \\
\end{eqnarray*}
 Then 
\begin{eqnarray*} 
\makebox[0ex][l]{$\displaystyle 
 \Big(\frac{\partial^2 }{\partial t^2} + \Omega^2 \Big) 
 \Big(Q_d(t)
  + D_{01}{2\gamma} \int_0^t \gamma_0 Q_d(\tau) d\tau \Big) 
  $}
\\&=&
 D_{01}
 \Bigg(
 {2\gamma} \int_0^t  
 \gamma_0 
 \Big(\Omega^2 Q(\tau)-\Omega^2 u_{01}(\tau,x_0)\Big) d\tau 
 +
 {2\gamma}\frac{\partial }{\partial t} \gamma_0 q(t) \Big|_{t=0} 
 - 
 \frac{\partial^2 }{\partial t^2}
 \Big({2\gamma} \int_0^t \gamma_0 u_{01}(\tau,x_0) d\tau \Big)
 \\&&{} \qquad \qquad
 + {2\gamma} \int_0^t \gamma_0 f_0(\tau) d\tau
 \Bigg)
 \\
\end{eqnarray*}
 Then 
\begin{eqnarray*} 
\makebox[0ex][l]{$\displaystyle 
 \Big(\frac{\partial^2 }{\partial t^2} + \Omega^2 \Big) 
 Q_d(t)
  + 
 D_{01}\Big(\frac{\partial^2 }{\partial t^2} + \Omega^2 \Big) 
 {2\gamma} \int_0^t \gamma_0 Q_d(\tau) d\tau 
  $}
\\&=&
 D_{01}
 \Bigg(
 {2\gamma} \int_0^t  
 \gamma_0 
 \Big(\Omega^2 Q(\tau)-\Omega^2 u_{01}(\tau,x_0)\Big) d\tau 
 +
 {2\gamma}\frac{\partial }{\partial t} \gamma_0 q(t) \Big|_{t=0} 
 - 
 \frac{\partial^2 }{\partial t^2}
 \Big({2\gamma} \int_0^t \gamma_0 u_{01}(\tau,x_0) d\tau \Big)
 \\&&{} \qquad \qquad
 + {2\gamma} \int_0^t \gamma_0 f_0(\tau) d\tau
 \Bigg)
 \\
\end{eqnarray*}
 Use now that  
$$
Q(t)-u_{01}(t,x_0) =: Q_d(t)
$$
 Then 
\begin{eqnarray*} 
\makebox[10ex][l]{$\displaystyle 
 \Big(\frac{\partial^2 }{\partial t^2} + \Omega^2 \Big) 
 Q_d(t)
  + 
 D_{01}\Big(\frac{\partial^2 }{\partial t^2} + \Omega^2 \Big) 
 {2\gamma} \int_0^t \gamma_0 Q_d(\tau) d\tau 
  $}
\\&=&
 D_{01}
 \Bigg(
 {2\gamma} \int_0^t  
 \gamma_0 
 \Omega^2 Q_d(\tau) d\tau 
 +
 {2\gamma}\frac{\partial }{\partial t} \gamma_0 q(t) \Big|_{t=0} 
 - 
 \frac{\partial^2 }{\partial t^2}
 \Big({2\gamma} \int_0^t \gamma_0 u_{01}(\tau,x_0) d\tau \Big)
 \\&&{} \qquad \qquad
 + {2\gamma} \int_0^t \gamma_0 f_0(\tau) d\tau
 \Bigg)
 \\
\end{eqnarray*}
 Then 
\begin{eqnarray*} 
\makebox[10ex][l]{$\displaystyle 
 \Big(\frac{\partial^2 }{\partial t^2} + \Omega^2 \Big) 
 Q_d(t)
  + 
  D_{01}
  \frac{\partial^2 }{\partial t^2}
  {2\gamma} \int_0^t \gamma_0 Q_d(\tau) d\tau 
  $}
\\&=&
 D_{01}
 \Bigg(
 {2\gamma}\frac{\partial }{\partial t} \gamma_0 q(t) \Big|_{t=0} 
 - 
 \frac{\partial^2 }{\partial t^2}
 \Big({2\gamma} \int_0^t \gamma_0 u_{01}(\tau,x_0) d\tau \Big)
 + {2\gamma} \int_0^t \gamma_0 f_0(\tau) d\tau
 \Bigg)
 \\
\end{eqnarray*}
 Then, finally,  
\medskip 

\hspace{\fill}
\fbox{
\hspace{\fill}
\parbox{1.01\textwidth}{
\begin{eqnarray*} 
\makebox[18ex][l]{$\displaystyle 
 \Big(\frac{\partial^2 }{\partial t^2}
 +D_{01}{2\gamma}\frac{\partial }{\partial t}\gamma_0 
 + \Omega^2 \Big) 
 Q_d(t)
  $}
\\&=&
 D_{01}
 \Bigg(
 {2\gamma}\frac{\partial \gamma_0 q(t)}{\partial t}\Big|_{t=0} 
 - 
 {2\gamma}\frac{\partial \gamma_0 u_{01}(t,x_0)}{\partial t}
 + {2\gamma} \int_0^t \gamma_0 f_0(\tau) d\tau
 \Bigg)
 \\
\end{eqnarray*}
\medskip 
\hspace{\fill}
\begin{eqnarray*}
 \mbox{ where }
 Q_d(t) := Q(t)-u_{01}(t,x_0)
 &,&
 D_{01} := \Big(\alpha_0 + \alpha_1 \frac{\partial}{\partial t} \Big)
 \\
\end{eqnarray*}
}
\hspace{\fill}
}
\hspace{\fill}

\medskip 

\noindent 
 On the surface, this equation appears to be a second order {\bf ordinary } 
 differential equation. It is not exactly the case. 
 We may not arbitrary take the initial data
 for
$Q(t)$
 because 
\begin{eqnarray*}
 Q(t)
 &=&
 \Big(\alpha_0 + \alpha_1 \frac{\partial}{\partial t} \Big)
 \Bigg(
 -{2\gamma}
 \int_0^t 
 \Big(\gamma_0 (Q(\tau)-q(\tau))\Big)d\tau
 + u_{01}(t,x_0) \Bigg)  
 \\
\end{eqnarray*}
 and  
\begin{eqnarray*}
 \frac{\partial Q(t)}{\partial t}
 &=&
 \Big(\alpha_0 + \alpha_1 \frac{\partial}{\partial t} \Big)
 \Bigg(
 -{2\gamma}
 \Big(\gamma_0 (Q(t)-q(t))\Big)
 + u_{01}(t,x_0) \Bigg)  
 \,,
\\&&
 \quad Q(0) 
 =
  \Big(\alpha_0 + \alpha_1 \frac{\partial}{\partial t} \Big)
 u_{01}(t,x_0)\Bigg|_{t=0}
 + \alpha_1 
 \Bigg(
 -{2\gamma}
 \Big(\gamma_0 (Q(0)-q(0))\Big)
 \Bigg)
 \\
\end{eqnarray*}

 The proper ones are these:

\begin{eqnarray*}
 \frac{\partial Q(t)}{\partial t}\Bigg|_{t=0}
 &=&
 \Big(\alpha_0 + \alpha_1 \frac{\partial}{\partial t} \Big)
 \Bigg(
 -{2\gamma}
 \Big(\gamma_0 (Q(t)-q(t))\Big)
 + u_{01}(t,x_0) \Bigg)\Bigg|_{t=0}  
 \,,
\\&&
 \quad Q(0) 
 =
  \Big(\alpha_0 + \alpha_1 \frac{\partial}{\partial t} \Big)
 u_{01}(t,x_0)\Bigg|_{t=0}
 + \alpha_1 
 \Bigg(
 -{2\gamma}
 \Big(\gamma_0 (Q(0)-q(0))\Big)
 \Bigg)
 \\
\end{eqnarray*}

%
%

 Thus, 
 we have already obtained two simple `ordinary' differential equations for 
$$
  q(t) \,,\quad Q(t) 
$$ 
 If we have found these quantities, we can try to find 
$u(t,x)$
 by
$$
\parbox{\textwidth}{
\begin{eqnarray*}
 u(t,x) 
 &=&
 \left\{ 
 \begin{array}{ccl}
 \displaystyle
 -{2\gamma}
 \int_0^{t-|x-x_0|/c} 
 \Big(F_{src}(\tau,q,Q)\Big)d\tau &,& \mbox{ if } 0 \leq t-|x-x_0|/c \\ 
                 0              &,& \mbox{ if } t-|x-x_0|/c < 0 \leq t \\ 
 \end{array}
 \right\} 
 \\[\medskipamount]\\
 &&{}\qquad + u_{01}(t,x) 
 \\[\smallskipamount] \\ 
\end{eqnarray*}
}
$$


\newpage 
\section{Appendix B. Remark. Complete Reflection }

 Suppose, we DO have a situation where 
$Q(t) \equiv 0$
 . 
 In that case 
$$
 \Big(\alpha_0 + \alpha_1 \frac{\partial}{\partial t} \Big)
 \Bigg(
 -{2\gamma}
 \int_0^t 
 \Big(F_{src}(\tau,q,Q)\Big)d\tau
 + u_{01}(t,x_0) \Bigg) =0  
$$
 and then 
$$
 -{2\gamma}
 \int_0^t 
 \Big(F_{src}(\tau,q,Q)\Big)d\tau
 + u_{01}(t,x_0) 
  = const_0 e^{-\alpha_{01} t} 
$$
 for suitable constants
$const_0, \lambda_0$
 . 
 Thus, we have 
\begin{eqnarray*}
 u(t,x) 
 &=&
 const_0 e^{-\alpha_{01} (t-|x-x_0|/c)}
\\&&{}
 - u_{01}(t-|x-x_0|/c,x_0) + u_{01}(t,x_0) 
\\&&{}
 \qquad\qquad    
 \mbox{ if } 0 \leq t-|x-x_0|/c
\end{eqnarray*}
 and, of course,  
\begin{eqnarray*}
 u(t,x) 
 &=&
 \begin{array}{ccl}
  u_{01}(t,x) &,& \mbox{ if } t-|x-x_0|/c < 0 \leq t \\ 
 \end{array}
\end{eqnarray*}

 Suppose in addition, that 
$u_{01}(t,x)$
 is of the form 
$u_{+}(t+x/c)$
 , i.e., 
$u_{01}(t,x)$
 is a "wave moving from right to left".
 Then 
\begin{eqnarray*}
 u(t,x) 
 &=&
 const_0 e^{-\alpha_{01} (t-|x-x_0|/c)}
\\&&{} 
 - u_{+}(t-|x-x_0|/c+x_0/c) + u_{+}(t+x/c) 
\\&&{}
 \qquad\qquad    
 \mbox{ if } 0 \leq t-|x-x_0|/c
\end{eqnarray*}
 Hence 
\begin{eqnarray*}
 u(t,x) 
&=&
\left\{
\begin{array}{lcl}
 const_0 e^{-\alpha_{01} (t-|x-x_0|/c)} 
\\ 
 const_0 e^{-\alpha_{01} (t-|x-x_0|/c)} 
 - u_{+}(t-x/c+2x_0/c) + u_{+}(t+x/c)
\\ 
 u_{+}(t+x/c) 
\end{array}
\right\}
\end{eqnarray*}
 if 
$$
\left\{
\begin{array}{ccl}
 0 \leq t-|x-x_0|/c \,,\, x \leq x_0 
\\ 
 0 \leq t-|x-x_0|/c \,,\, x \geq x_0 
\\ 
 t-|x-x_0|/c < 0 \leq t 
\end{array}
\right\}
 \,,
$$
 respectively. It is because 
\begin{eqnarray*}
 - u_{+}(t-|x-x_0|/c+x_0/c) + u_{+}(t+x/c) = 0 
&\mbox{ if }& 
 x \leq x_0 
\\
  -u_{+}(t-|x-x_0|/c+x_0/c) = -u_{+}(t-x/c+2x_0/c) 
&\mbox{ if }& 
 x \geq x_0 
 \,.  
\end{eqnarray*} 
 In this situation we can say, that the incident wave, 
$u_{01}(t,x)=u_{+}(t+x/c)$, 
 is {\bf completely reflected, rejected, } by the oscillator.

\newpage 
 Some interesting details:

 For the 
$F_{src}, f_{compl}$
 we wish to discuss, 
 the conditions 
$$
 Q = 0 \,,\, f_0 = 0 
$$
 mean that 
$$
  \frac{\partial^2 q(t)}{\partial t^2}
  =-\Omega^2 q(t) \,.
$$
 That is,
$$
 q(t) = A_s\sin(\Omega t) + A_c\cos(\Omega t)
$$ 
 where 
$A_s,A_c$
 are suitable constants. 
 Therefore, the complete reflection occurs
\footnote{
 recall, 
$F_{src}(\tau,q,Q) 
 = const_1 q(\tau) + const_2 \frac{\partial }{\partial \tau}q(\tau)
   + const_3 Q(\tau)
$}
, 
 where 
$$
 u_{01}(t,x_0)
  = \tilde A_s\sin(\Omega t) + \tilde A_c\cos(\Omega t) + \tilde A_0 
    + const_0 e^{-\alpha_{01} t} 
$$
 for suitable constants 
$\tilde A_s , \tilde A_c , \tilde A_0$
 . 
 Thus, we have seen: 
 The complete reflection occurs only where 
$ u_{01}(t,x_0) $
 is trigonometric up to 
$const_0 e^{-\alpha_{01} t}$; 
 moreover, ---only where the spectrum of the 
{\sc trigonometric part of } 
$ u_{01}(t,x_0) $
 contains only one non-zero (and real) frequence;
 moreover, ---only where this 
{\sc frequence coincides with the eigenfrequence of the oscillator.}


\newpage 

\bibliographystyle{unsrt}

\end{document}